# Establishing Trust in Online Advertising With Signed Transactions

ANTONIO PASTOR[1], RUBÉN CUEVAS[1,2], ÁNGEL CUEVAS[1,2], AND ARTURO AZCORRA[1,3], (Senior Member, IEEE)
[1]Telematic Engineering Department, Universidad Carlos III de Madrid, 28918 Leganés, Spain
[2]UC3M-Santander Big Data Institute, 28903 Getafe, Spain
[3]IMDEA Networks Institute, 28911 Leganés, Spain

Corresponding author: Antonio Pastor (anpastor@it.uc3m.es)

The research leading to these results has received funding from: the European Union's Horizon 2020 innovation action programme under grant agreement No 871370 (PIMCITY project) and grant agreement 78674 (SMOOTH Project); the Ministerio de Economía, Industria y Competitividad, Spain, and the European Social Fund (EU), under the Ramón y Cajal programme (grant RyC-2015-17732), project TEC2016-76795-C6-3-R, and the FPI grant BES-2017-080053; and the TAPTAP Digital-UC3M Chair in Advanced AI and Data Science applied to Advertising and Marketing.

**ABSTRACT** Programmatic advertising operates one of the most sophisticated and efficient service platforms on the Internet. However, the complexity of this ecosystem is a direct cause of one of the most important problems in online advertising, the lack of transparency. This lack of transparency enables subsequent problems such as advertising fraud, which causes billions of dollars in losses. In this paper we propose *Ads.chain*, a technological solution to the lack-of-transparency problem in programmatic advertising. *Ads.chain* extends the current effort of the Internet Advertising Bureau (IAB) in providing traceability in online advertising through the *Ads.txt* and *Ads.cert* solutions, addressing the limitations of these techniques. *Ads.chain* is (to the best of the authors' knowledge) the first solution that provides end-to-end cryptographic traceability at the ad transaction level. It is a communication protocol that can be seamlessly embedded into ad-tags and the OpenRTB protocol, the de-facto standards for communications in online advertising, allowing an incremental adoption by the industry. We have implemented *Ads.chain* and made the code publicly available. We assess the performance of *Ads.chain* through a thorough analysis in a lab environment that emulates a real ad delivery process at real-life throughputs. The obtained results show that *Ads.chain* can be implemented with limited impact on the hardware resources and marginal delay increments at the publishers lower than 0.20 milliseconds per ad space on webpages and 2.6 milliseconds at the programmatic advertising platforms. These results confirm that *Ads.chain*'s impact on the user experience and the overall operation of the programmatic ad delivery process can be considered negligible.

## I. INTRODUCTION

Online advertising is a multi-billion dollar business. The Internet Advertising Bureau (IAB) reported that the revenue generated by online advertising was $124 B in 2019, with an inter-annual growth rate of 16 % [1]. Besides, online advertising is the main revenue source of some of the most important Internet companies, such as Facebook [2] or Google [3], which are fundamental contributors to Internet innovation.

Programmatic advertising operates one of the most sophisticated and efficient service platforms on the Internet, which allows to deliver tailored ads based on tens of parameters (e.g., interests and online behavior of users, the context of the website/mobile app) through a real-time auction process. The overall process occurs in the order of hundreds of milliseconds and runs on top of a very complex ecosystem depicted in Fig. 1. In particular, the process of delivering an ad from an advertiser to a website or mobile app[1] involves several players, such as Demand-Side Platforms, Ad Exchanges, Supply-Side Platforms, and Ad Networks. The revenue generated by the impression of the advertiser's ad is then split between the website and the involved intermediaries.

The complexity of this ecosystem is a direct cause of one of the most important problems in online advertising, the lack of transparency. This lack of transparency enables

The associate editor coordinating the review of this manuscript and approving it for publication was Kuo-Hui Yeh.

---

[1]For clarity, we refer only to websites where websites and mobile apps are equivalent. Differences are contemplated explicitly.

  

subsequent problems such as advertising fraud–which attracts between 5 % and 19 % of the overall online advertising revenue [4], [5]–or misreporting of ad campaign information to advertisers [6], [7]. The online advertising industry has reacted to the *accusation* of lack of transparency creating auditing companies referred to as verifiers. However, in practice, these verifiers also use opaque auditing techniques that do not help to solve the problem [4], [5], [7], [8].

In this paper we propose *Ads.chain*, a technological solution to the lack-of-transparency problem. In particular, we define a communication protocol to provide end-to-end cryptographic traceability at the ad transaction level. In *Ads.chain*, every intermediary involved in the delivery process of an ad has to include a digital signature in the messages passed to its buy-side partner. The signature certifies the integrity and non-repudiability of the parameters containing relevant information. Therefore, each ad transaction produces a chain of digital signatures including the identity of each of the involved intermediaries and their actions. These chains of signatures provide guarantees of full transparency since any illegal or inappropriate action, as well as its perpetrator, can be identified by auditing the chains. Besides, the design of the protocol as a chain of signatures allows an incremental adoption by the industry.

*Ads.chain* can be seamlessly embedded into ad-tags and the OpenRTB protocol, the de-facto standards for communications between intermediaries in the online advertising ecosystem. Moreover, it leverages the existing Public Key Infrastructure (PKI) used for HTTPS communications to emit the public key certificates for the validation of the digital signatures. Hence, the protocol is readily implementable in the current ecosystem without requiring any modification, lowering the entry barrier for its adoption significantly.

We have implemented the protocol and made the code available through the following GitHub repositories [9], [10]. We assess the performance of *Ads.chain* through a thorough analysis in a lab environment that emulates a real ad delivery process at real-life throughputs. The obtained results show that *Ads.chain* can be implemented with limited impact on the hardware resources and marginal delay increments at the publishers lower than 0.20 ms per ad space on webpages and 2.6 ms at the programmatic advertising platforms.

The rest of the paper is organized as follows. Section II describes the operation of the programmatic online advertising ecosystem and a high-level introduction to digital signatures. Section III introduces *Ads.chain*, our proposed protocol to provide end-to-end traceability to individual ad transactions in the online advertising ecosystem. Section IV details the implementation of the proposed protocol and the lab environment in which we test its performance as described in Section V. Finally, Section VI concludes the paper.

## II. BACKGROUND

In this section, we provide an overview of the operation of the online advertising ecosystem and relevant details of ad-tag calls and OpenRTB messages, where our traceability protocol is embedded. Besides, we summarize some of the proven consequences of the lack of transparency and discuss some of the proposals from the industry that partially address this problem. Finally, we also provide a high-level introduction to digital signatures, which enable our protocol.

### A. ONLINE ADVERTISING OVERVIEW

In its inception, online advertising mimicked the scheme used in traditional media advertising, where advertisers and publishers (i.e., owners of websites and mobile apps) closed deals to show advertisers' ads on publishers' websites through a direct agreement or involving just one intermediary. In recent years, the model has rapidly evolved to the known as *programmatic* advertising ecosystem where ads are traded through a complex and heterogeneous set of automated platforms, often individually and in real-time. These platforms communicate among them to serve a suitable ad for a predefined ad space on a webpage. The messaging mechanisms and protocols these entities use to communicate with each other are fairly standardized and described later in this section.

#### 1) PROGRAMMATIC ADVERTISING OPERATION
When a user visits a website, the HTML document of the webpage is requested from the web server of *publisher.com*. This HTML document contains an ad-tag for every ad space the publisher allocates on the page. Ad-tag is the term used for the HTML code snippets containing the URLs to retrieve ad-related content [11]. Each ad-tag contains a URL pointing to the publisher's ad server to which the user's browser performs an HTTP request starting the ad serving process. The URL of the ad-tag contains information about the ad space.

Upon the reception of a request from the user's browser, the publisher's ad server may enrich the request with information about the user profile using proprietary information or requesting it to a Data Management Platform (DMP).[2] If the ad server finds a pre-configured ad campaign suitable for the user's profile, the user's browser retrieves the pre-configured ad from the advertiser's ad server. These pre-configured campaigns typically correspond to private deals between advertisers and publishers. In case there are no pre-configured campaigns for the user's profile, the ad server forwards the ad request to a Sell-Side Platform (SSP) or ad network. These traders try to sell the ad space through private exchanges where only a selected group of buyers have access.

If the ad request is still not sold through these private channels, the publisher's ad server places the ad request in the open market. This process is usually done through an SSP that forwards the ad request to an Ad Exchange (AdX) [12]. The AdX launches a real-time open auction sending bid request messages to Demand-Side Platforms (DSPs). This bid request message includes information about the ad space (e.g., type, size, and location in the webpage), the domain (i.e., the website or mobile app), and the user's profile and device. The DSPs are entities where the advertisers' ad campaigns are pre-configured. Upon the reception of the bid request, a DSP checks whether the parameters included in the bid request match any of the pre-configured campaigns. If so, they respond to the bid request with a bid response that includes the bidding price and the ad-tag with the ad's URL. Then, once the bidding time has expired, the AdX runs the auction and: (1) informs of the result to the DSPs with win and loss notice messages; (2) forwards the ad's URL through the inverse chain of communication from the AdX to the user's browser. Finally, the advertiser's ad server, which is usually hosted at its DSP's ad server, receives the request for the ad from the user's browser and accounts for the impression as performed.

---

[2]Note that DMPs can be queried from any other intermediary entity in the ad delivery process.

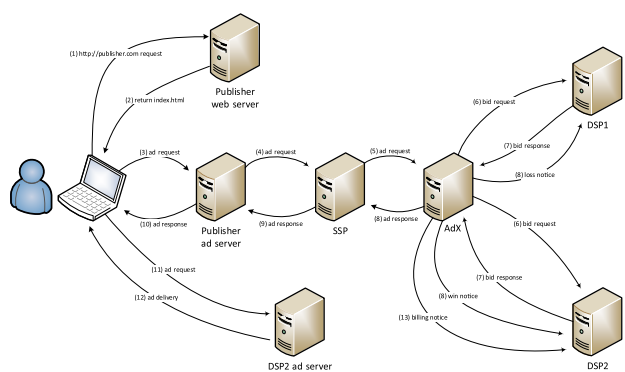

**FIGURE 1.** Overview of the programmatic advertising ecosystem. The arrows represent the flow of messages in the ad delivery process.

Fig. 1 graphically depicts the process described above. The overall process takes in practice less than a second, from which the auction process is restricted to less than 300 ms in most AdX.[3]

#### 2) MESSAGE FORMATS AND COMMUNICATION PROTOCOLS

The complex procedure described above relies on the exchange of different types of messages. We can differentiate two clear parts in the overall programmatic process. On the one hand, the sell-side involves all entities participating in the process until the communication reaches the AdX: the publisher, the publisher's ad server, and the SSP. On the other hand, the buy-side is formed by the DSPs. Finally, the AdX is the entity communicating the sell and buy sides.

The communication between AdXs and DSPs in the buy-side uses the Open Real-Time Bidding (OpenRTB) protocol, a standard defined by the IAB and adopted by the industry. The OpenRTB defines the format and order of messages exchanged between AdXs and DSPs (bid request, bid response, win/loss notice, and billing notice). OpenRTB uses HTTP as the communication protocol and JSON (JavaScript Object Notation) format for data serialization. The latest operational version is v3.0. It was released in

[3]https://developers.google.com/authorized-buyers/rtb/start

November 2018 and includes a beta version of *Ads.cert*, a mechanism to provide signed bid requests, which is one of the basic components we leverage in our solution to create an end-to-end chain of signatures per ad transaction.

The sell-side entities rely on ad-tags as mean to communicate with each other. The response to an ad-tag call (a request to the URL of an ad-tag) can be another ad-tag or the final advertisement. The structure of ad-tags may vary depending on its function. They may include JavaScript code to perform dynamic tasks at rendering time or even a *no-script* section for the browsers with JavaScript disabled. The information about the ad impression (such as iframe size, user's profile, or the winning price) is embedded in the URL's query string, the URL part after the question mark symbol (?). The query string parameters are separated using an ampersand symbol (&) and have a key-value format using an equal sign (=) between the keys and their respective values. Fig. 2 shows an example of an ad-tag.

```
<script
  src="https://ssp.com/ttj?id=123e45b7"
  type="text/javascript">
</script>
```

**FIGURE 2.** Example of an ad-tag defining an ad space on a web page.

### B. LACK OF TRANSPARENCY AND FRAUD IN ONLINE ADVERTISING

#### 1) PROBLEM DESCRIPTION

The online advertising industry has managed to develop a very efficient ecosystem that is able to deliver tailored ads involving a real-time auction process in a few hundreds of milliseconds. However, this technology development lacks appropriate, objective, and transparent auditing mechanisms. There is no way to check the validity or veracity of the parameters that an entity A passes to an entity B, either through ad-tags or OpenRTB messages. Moreover, advertisers are left out of the process, and what they receive are processed reports summarizing the performance of their campaigns, which have been reported to be inaccurate [6]. Besides, this lack of transparency enables ad fraud, one of the most important problems of online advertising.

There are different reported forms of ad fraud: from basic attacks using bots to visit websites where ads are shown [6], [13] and even clicked [14], to more sophisticated attacks using malicious software–referred to as *adware*–that performs hidden visits to websites from a user's browser [15]. Given the lack of transparency, the fraud problem is not isolated to artificial traffic to untrustworthy publishers. Recent reports document high scale cases of counterfeit inventory fraud. In this type of attack, fraudsters take advantage of the impossibility to validate the veracity of the information included in ad-tags or OpenRTB messages. Domain spoofing is a well-known attack to introduce counterfeit inventory in the programmatic ecosystem [16]. In particular, fraudsters launch fraudulent ad requests from instrumented browsers claiming to come from popular domains, referred to as *premium sites*, where ad spaces are more expensive.

#### 2) STATE-OF-THE-ART SOLUTIONS

**- Proprietary Solutions:** Pushed by advertisers' concerns about the lack of transparency, several independent companies referred to as verifiers have appeared in the last years, e.g., IAS [17], White Ops [18], DoubleVerify [19]. These companies use ad-tags embedded in publishers' websites, containers of ads such as iFrames, or the ad creativity to monitor the delivery process of individual ad impressions. However, these companies operate in an opaque manner. They use proprietary technology, which has not been validated and does not address the problem of lack of transparency in a holistic and efficient manner. Indeed, the effectiveness of their technology has been questioned by research studies [7], [8], suggesting that solving the lack of transparency and fraud problems using opaque auditing techniques is not an appropriate approach.

**- IAB promoted standard solutions:** The IAB Tech Lab has proposed open source standards to address the lack of transparency and fraud problems: *Ads.txt* and *Ads.cert*. However, these solutions were mainly driven by the recently discovered domain spoofing fraud attack. Hence, as the existing proprietary solutions, IAB standards are not designed to solve the lack of transparency in a holistic manner.

The Authorized Digital Sellers *Ads.txt* specification, launched in 2017, consists of a plain text file where publishers publicly declare the traders (e.g., ad networks, SSPs, and exchanges) with which they operate. Hence, any player can check whether the ad request comes from a valid trader. This ad-hoc solution has obvious limitations. There is a not negligible portion of publishers not adopting it. Moreover, *Ads.txt* imposes to blindly trust authorized sellers [20] and authorized resellers that could have received the requests through an unauthorized source [16]. Given the limitations of *Ads.txt*, the IAB launched the *Ads.cert* specification, whose beta version is included in OpenRTB 3.0. *Ads.cert* defines a standardized mechanism by which the publishers can sign the ad requests using public-key cryptography to provide proof of their identity. Although this step goes in the right direction to address the counterfeit inventory, it is not an end-to-end solution to provide full transparency to ad transactions and has limitations, as recognized by the IAB in Section 6 of the same specification.[4] For instance, the current definition of *Ads.cert* introduces a new vulnerability: it allows a malevolent platform in the selling chain to replicate signed ad requests originated at a compromised user's browser. The replicated ads can be sold to different buyers or even to the same DSP if appropriate sanity checks in the received inventory are not performed.

**- *Ads.chain* vs. state-of-the-art solutions:**

Extending these initial efforts by the IAB, we define *Ads.chain*, a cryptography-based solution that, for the first time, provides end-to-end traceability in the ad delivery process at the level of individual transactions. This guarantees transparency, accountability, and non-repudiability for every ad transaction.

In particular, compared to existing IAB standards, *Ads.chain* provides a more dynamic solution than *Ads.txt* to mitigate the possibility of introducing counterfeit inventory and extends the signatures defined in *Ads.cert* to the complete chain of custody of each ad transaction to avoid vulnerabilities, such as the replication of ad requests.

Finally, existing proprietary solutions do not offer end-to-end traceability and do not rely on cryptographic solutions. Most of them are sophisticated measurement-based solutions to assess the performance of ad campaigns and identify fraudulent activity. However, they are unable to provide fundamental functionalities such as accountability and non-repudiability. Furthermore, these proprietary solutions are, by definition, non-transparent, whereas *Ads.chain* is open source.

### C. DIGITAL SIGNATURES

Digital signatures provide cryptographic proof of data integrity, data origin authentication, and non-repudiation. The schemes in use on today's Internet to produce digital signatures are based on asymmetric cryptography, also known as public-key cryptography, as each entity's key has two parts, a public key that is distributed to others and a private key that is kept as a secret [21].

Asymmetric cryptography provides the capability of decrypting with the public key a message encrypted with the private key and vice-versa, but not with the same key part. This particularity of public key cryptography is used to produce digital signatures by encrypting a checksum of a message, produced using a hash function, with the private key. The encrypted checksum is the signature and is transmitted with the message. Any other entity can then verify the signature using the signer entity's public key to decrypt the signature and computing the checksum of the message with the same hash function used in the signing process. If the decrypted checksum from the signature matches the checksum computed from the received message, it can be assumed that the message has not been modified and was signed with the paired private key (i.e., was generated by the owner of the public and private key pair).

#### 1) PUBLIC KEY INFRASTRUCTURE AND DIGITAL CERTIFICATES

To provide non-repudiation on digital signatures, the ownership of a public key has to be provable. Similarly, an entity may need to revoke a public key to no longer consider it valid for validating signatures; for instance, if the paired private key might have been compromised. The X.509 Public Key Infrastructure (PKI) for the Internet enables the distribution of public keys providing such services using digital certificates signed by Certification Authorities (CAs) [22], [23].

---

[4]https://github.com/InteractiveAdvertisingBureau/openrtb/blob/master/ads.cert:%20Signed%20Bid%20Requests%201.0%20BETA.md#6-limitations-and-abuse-vectors-

A CA is an entity that validates the identity of other entities and issues digital certificates for them. A digital certificate is an electronic document that binds a public key to an entity including additional information, such as the key's validity period. This document is signed by the issuer CA to provide trust in it. A domain validated certificate can be considered sufficient to accept a DNS domain's public key as valid. For issuing a domain validated certificate, the CA checks the administrative control of the fully qualified domain name [24].

A software trusts a CA if one of two conditions is met. The first one is that the root certificate, a self-signed certificate of the CA, is stored on the certificate store used by the software. The second condition is that the CA certificate is issued by a trusted CA or the chain of certificates identifying the CAs (called validation path) leads to a trusted CA and the CA delegation conditions are valid. The X.509 PKI defines delegation mechanisms for CAs, and the terms under which a certificate not directly signed by a root CA can be considered valid [22], [25]. To be accepted as a trusted root CA by major browsers [26], [27], a CA can only accept the algorithms and minimum key sizes specified in the Baseline Requirements of the CA/Browser Forum [28], an industry consortium of browser vendors and public trust centers.

### 2) DIGITAL SIGNATURE ALGORITHMS

The security of asymmetric cryptography schemes is based on the intractability of their underlying mathematical problem. Specifically, the integer factorization problem for the RSA digital signature algorithm or the elliptic curve discrete logarithm problem for ECDSA (Elliptic Curve Digital Signature Algorithm) [21]. Following the Baseline Requirements of the CA/Browser Forum and *Ads.cert*, we analyze the performance for RSA using a key size of 2048 bits and for ECDSA using the curve NIST P-256. The RSA key size of 2048 bits is the minimum accepted key size for this algorithm. Hence, it shows the maximum performance achievable with this algorithm as longer keys would translate into slower operation times. Among the accepted curves for ECDSA, we use the curve NIST P-256 since it is the one proposed for *Ads.cert*. In both cases, we use the hash function SHA-256 (as in *Ads.cert*) to compute the message digest.

The level of security is determined by the weakest of the two algorithms (signature algorithm and hash function) used to produce a digital signature. RSA keys of 2048 bits provide a security strength of 112 bits, whereas the security strength of both the curve NIST P-256 and the SHA-256 hash algorithm is 128 bits [29]. In terms of performance, comparative studies show that RSA-2048 is faster than ECDSA P-256 for signature verification [30], whereas ECDSA P-256 can outperform RSA producing a signature [31]. However, optimizations on the code and processor architectures may impact the final performance. Therefore, we conduct a performance evaluation (using OpenSSL version 1.1.1g) to measure the marginal delay that *Ads.chain* operations can introduce on ad transactions.

## III. *Ads.chain* PROTOCOL DESIGN

In this section we describe in detail *Ads.chain*. We start identifying the requirements the protocol has to meet to achieve its purpose (end-to-end traceability of ad transactions) while being implementable in the current programmatic ecosystem. Then, we describe the protocol, and finally, we describe how to seamlessly integrate it into the current online advertising ecosystem.

### A. PROTOCOL REQUIREMENTS

- **Unequivocal custody**: in an ad delivery process, only one player has the right to re-sell the ad space at a time, following the scheme described in Fig. 1. In other words, only one player has the custody of the ad space at a given moment. In the current ecosystem, a malicious player may declare to own the custody of an ad space, and there is not an easy way to prove if it is true or false. Hence, the defined protocol must guarantee the unequivocal custody principle by which it is verifiable if a player owns the custody of an ad space.
- **Non-repudiability**: any action taken by a player should be undeniable. This property is referred to as non-repudiability in the security discipline.
- **Low latency**: the overall ad delivery process takes hundreds of milliseconds in programmatic advertising. Therefore, the protocol must incur delays in the order of a few ms to have a minimal impact on the overall delivery process. On the other hand, the impact on the overall page loading time should be likewise small to avoid affecting the end-user experience.
- **Scalability**: the online advertising ecosystem delivers around a trillion ads every day. The protocol must be able to operate at this scale.
- **Seamless integration**: the protocol must allow its integration as part of the existing protocols and methods in online advertising without the need to modify them. In particular, it must be implementable in ad-tags and as part of the OpenRTB protocol, the two methods used for communication in the sell- and buy-side of the online advertising ecosystem, respectively.
- **Online and offline auditing**: the protocol must allow two types of auditing operations. On the one hand, online auditing enables an entity to audit the validity of a received ad transaction in real-time. On the other hand, each ad transaction must create a log that can be audited so that misbehaving players can be identified at any moment in the future.

### B. PROTOCOL OVERVIEW

In essence, an ad transaction can be defined by a chain of individual actions taken by the involved players in the ad delivery process (See Fig. 1). These actions are, in many cases, subject to the terms of a contract signed between two entities.

We propose to generate a digital chain that records the actions of every player involved in an ad transaction. Conceptually, the chain is formed by blocks. Each block is inserted in the chain by a player participating in the ad delivery process and summarizes the most relevant parameters associated with the actions taken by the player. Moreover, the block is signed with a private key that unequivocally identifies the player. Finally, a block is linked with the previous block to form the chain.

Following this simple protocol, the actions of the first player in the ad delivery process (i.e., the publisher) are recorded in the first block of the chain. This first block includes: 1) An universally unique identifier of the ad transaction. 2) Information identifying unequivocally the player to which the custody of the ad transaction is assigned so that this player is the unique one with the rights to re-sell that ad space. This identifier is the player's domain name. 3) A foolproof identifier of the user to let the advertiser verify the final destination of the ad impression. This identifier is the IP address of the device requesting the ad. 4) Data fields, which are key-value pairs, where the actions of the publisher are registered. For instance, these may include the location of the ad space on the screen and the size of the ad space. Once all these data are compiled in the proper format, the publisher signs this block with its private key (for which the paired public key is publicly available in a digital certificate). Then, the block is generated and sent to the second player in the chain indirectly through the user browser.

The first action of this second player (e.g., an SSP) upon the reception of the first block is to verify the signature. If the signature is correct, it generates a second block. Otherwise, it rejects the ad transaction and informs about it to the publisher. This second block is simpler than the first block. It includes the signature of the first block creating the binding between blocks to form the chain. In addition, it includes the key-value data fields recording the relevant information associated with the actions taken by the SSP and the identity (domain name) of the third player to which the custody of the ad transaction is delegated. These data are signed with the private key of the SSP, and thus, the second block is created. The chain, now formed by two blocks, is sent to the third player.

If the ad delivery process involves n players, the associated chain has also n blocks. From the second to the last block, all have the same format described in the previous paragraph. The only differences among blocks 2 to n correspond to the data fields (key-value pairs), which may be different for different types of players (e.g., the data fields from an SSP and an AdX might be different). The first block is the only one having a different format since it includes the ad transaction ID and the IP address of the device, as described above. The last block of a chain is typically generated by the DSP winning the last auction of the ad transaction. Note that the advertiser winning the ad space associated with the transaction can verify the complete chain[5] to audit that no information has been tampered during the process. Moreover, the $i^{th}$ player in the chain can validate the blocks of players 1 to $i-1$. Hence, a malicious player in position $i$, which tries to modify previous blocks, can be easily identified by the advertiser or any player from position $i+1$ since the signature of the modified blocks will be incorrect.

In the previous paragraph, we are considering a distributed auditing scheme where advertisers take the responsibility of auditing its own ad transactions. Alternative auditing schemes can be defined, e.g., a centralized auditing entity that receives the chains and performs a central auditing process or an auditing entity defined by each publisher to validate the ad transactions associated with its websites. Note that it is up to the industry to choose the most appropriate auditing approach.

Finally, it is worth reviewing how this simple protocol meets the requirements defined above:

- unequivocal custody: only one player has the right to re-sell the ad space at each step of the process.
- non-repudiability: every action reported by a player is recorded and signed with its private key, for which the accompanying public key is distributed in a digital certificate. If a player takes inappropriate actions, it would be registered and can be proved later.
- low latency: entities executing our protocol need to perform and verify digital signatures in real-time. The execution time for these operations depends on several factors, such as the digital signature algorithm, the software implementation, and the hardware in which they are executed. We explore this specific aspect in detail in Section V. We show that *Ads.chain* meets the low latency requirement relying on 1) digital signature algorithms widely used in today's Internet communications, 2) open-source software, and 3) household-level commodity hardware.
- scalability: the protocol operates at the level of individual ad transactions, and its scalability may depend on implementation factors. However, we find that implementations achieving the low latency requirement also provide a great degree of scalability (See Section V).
- seamless integration: as we show in Section III-D, our protocol can be implemented with both ad-tags (used by sell-side entities) and OpenRTB (used by buy-side entities).
- online and offline auditing: our proposal allows a player that receives the custody of an ad transaction to audit the received chain in real-time. The player can then reject the transaction if there is any problem with it. Offline auditing is also possible using the chains associated with finalized transactions.

---

[5]We are assuming that a DSP will deliver to each advertiser the chains associated with its delivered ads.

### C. PROTOCOL DETAILS

In this subsection, we provide further technical details about the design of the protocol.

#### 1) UNIQUE IDENTIFIER OF AD TRANSACTION

The transaction identifier needs to be unique within each publisher's domain. Popular websites are typically served from a distributed infrastructure of servers–e.g., a Content Distribution Network (CDN)–and thus would require to generate concurrent identifiers from multiple servers. Therefore, they need a systematic scheme to generate identifiers at high throughputs without collisions.

In OpenRTB 3.0 and *Ads.cert*, the IAB mentions the need for having a unique transaction identifier to avoid replay attacks. However, it does not describe a format for it. We propose to use the Universal Unique Identifier (UUID) format described by RFC 4122 [32]. It is a well known and widely used standard for UUIDs, and there is available code to generate it in multiple programming languages.

A timestamp-based UUID is formed by 128 bits codifying three fields: a high-resolution timestamp (60 bits), a clock sequence (14 bits), and a node ID (48 bits). Popular domains need to serve ad spaces at a high rate. To meet this and future higher demands, we propose to provide a resolution of 1 ns. To this end, we borrow 7 bits from the clock sequence and assign them to the timestamp. That adjustment leaves another 7 bits for the clock sequence so that we can have up to 128 processes generating timestamps on a single server. With this format, the theoretical limit of the number of UUIDs per server is 128 billion per second. Similarly, we can codify timestamps up to the year 5623 if we use the UNIX epoch. Hence, it offers enough resolution and scalability to implement it even in the aforementioned distributed architectures such as CDNs, which may be serving tens of thousands of different domains.

In *Ads.chain* we use the string representation of a UUID–hexadecimal values of the 16 bytes (128 bits) separated with dashes after the fourth, sixth, eighth, and tenth byte–as the transaction UUID and refer to it as tUUID. The tUUID is generated by the publisher and tied to the ad transaction since the first signed block.

#### 2) BLOCKS CODIFICATION

A block in our protocol is formed by a set of data fields (key-value pairs) and the identity (domain name) of the next player in the chain, which is also represented in the format of a key-value pair. These data are certified by the digital signature of the player that generated them, which is also part of the block.

In order to include the block information in the ad transaction data, we only need to include three strings at every step: the *custody* field specifying the entity to which the sell is delegated, a *keys-string* that concatenates the keys of the fields included in the signature–separated with a special delimiter character–and the *signature* string codified in base64.

The data signed is a string with the values corresponding to the keys in the *keys-string*, in the same order. The signatures are performed over the SHA-256 hash digest of this string of values. Note that we do not need to include this string in the request as the values are already included in the request data, but we need the *keys-string* to be able to form it.

#### 3) HANDLING AUCTION PROCESSES

In an auction process, the custody of the transaction cannot be delegated until the process is concluded, and the winner of the auction (which will be the one receiving the custody delegation) is known.

Therefore, the entity launching the auction does not delegate the custody initially. Instead, it provides a temporary chain where its last block includes a temporary signature. This chain allows the participants in the auction to validate that the auction is run by the entity (usually an AdX) owning the custody of the ad transaction, as well as to check the information included in previous blocks.

When the auction is completed, the winner entity receives the OpenRTB billing notice message, including the final chain of blocks that delegates the custody of the ad transaction to the winner entity. The last block of this chain includes the domain name of the winner entity in the corresponding field.

If the auction corresponds to the last event of the ad delivery process, the winning entity, typically a DSP, generates the last block of the chain. This block signed by the DSP should include information regarding the advertiser, campaign id, and creativity associated with the ad delivered to the user.

#### 4) PUBLISHER SIGNATURE IN MOBILE APPS

*Ads.chain* is designed to work independently of ad transactions on websites or mobile apps. However, transactions in mobile applications start on the user device with a request generally to a sell-side platform instead of a publisher's server. In this case, the app creates the Ads.chain block and requests a trusted server of the app's publisher to sign it before sending it to the next intermediary in the custody chain. Note that many apps already interact with their backends, so adding this functionality is expected to require little development effort.

### D. SEAMLESS INTEGRATION

*Ads.chain* can be seamlessly integrated into ad-tags and OpenRTB, which are the de-facto standard communication techniques used in the sell and buy sides of the online advertising ecosystem, respectively. Both ad-tags and OpenRTB messages specify the parameters in key-value pairs. Hence, we can embed the block fields (i.e., *signature*, *custody*, and *keys-string*) while maintaining compatibility with current implementations. Specifically, in the ad-tag's URL, the parameters are appended to the query string as there is no hierarchical structure. Whereas, in OpenRTB objects, the fields should be included in the *Source* object as proposed in OpenRTB v3.0 for *Ads.cert* [33].

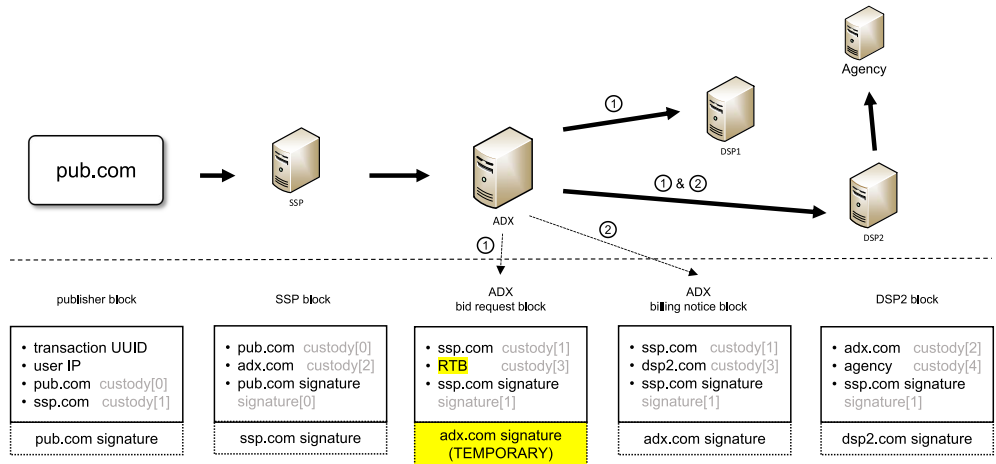

**FIGURE 3.** Representation of the *Ads.chain* block at each level of an ad transaction. The AdX block in the final chain is the block included the billing notice sent only to the DSP winning the auction. The bid request block is temporary for the RTB auction process and thus it is not received by the advertiser's agency.

Entities not implementing *Ads.chain* would only need to ignore the associated fields. However, these entities would generate a gap in the chain of custody since they do not generate a block in the chain. We conjecture that with the incremental adoption of our protocol, entities failing to implement it may be penalized in different manners. Some players may pay less for ad transactions having gaps in the trust chain due to the associated trust issues. Other players may directly reject ad transactions that do not implement *Ads.chain* end-to-end. These penalties may be an incentive for a faster adoption of the proposed protocol.

Finally, we propose to use for the public key certificates the existing X.509 Public Key Infrastructure (PKI) for the Internet that provides support to HTTPS communications. In order to distribute the public key certificates, we propose to use domain validated certificates of a specific subdomain of the signer entity–e.g., *ads.example.com* for the entity *example.com*–issued by existing CAs trusted by major browsers. This allows restricting the use of the key only for the purpose of *Ads.chain* following the best practices recommended by the US National Institute of Standards and Technology (NIST) [29]. Therefore, servers of this subdomain should not start the key exchange process on SSL connections but expose the certificate to the client. An alternative solution is to make the certificate accessible under a standard relative path, e.g., *example.com/ads-chain.crt*. This alternative solution is similar to the one proposed for *Ads.cert*; however, we emphasize on using digital certificates to ensure the non-repudiation of signatures.

### E. TRUST CHAIN EXAMPLE
Fig. 3 shows an example of how the final chain received by an advertiser looks like. Moreover, in the following link,[6] interested readers can access the format of the chain received by the different players of an ad transaction example involving a publisher web server, an SSP, an AdX, and a DSP winning the auction process and delivering the ad.

### F. Ads.chain VS. BLOCKCHAIN
Every ad transaction produces a chain of signed blocks. The demanding time constraints for delivering ads to users in real-time make impractical annotating these individual signed blocks of a chain as entries of a blockchain distributed ledger. Blockchain inspired solutions are more suitable for offline (not real-time) processes in the context of online advertising. For instance, they can be used for the verification of authenticity and uniqueness of the *Ads.chain* transaction chains.

## IV. Ads.chain IMPLEMENTATION AND LAB PROTOTYPE
To test the viability and performance of the proposed protocol, we have built a lab scenario with the main entities present in the delivery of ad transactions in programmatic advertising. We have also implemented an external library that offers all the code components required to implement *Ads.chain* by any of the platforms involved in an ad delivery process. Both the entities of the lab prototype and the library are implemented in C++, and their associated code is publicly available on GitHub [9], [10].

### A. LAB PROTOTYPE OF ONLINE ADVERTISING ECOSYSTEM
We have reproduced in our lab prototype a scenario similar to the one depicted in Fig. 1. In particular, it is formed by a publisher's website server, a sell-side entity acting as the publisher's SSP, an AdX connecting the sell and buy sides, and a DSP as the buy-side entity. There is also an ad server for serving the final ad to the user's browser upon the reception of the ad-tag of the auction winner.

We deploy each entity in independent instances in a private OpenStack with two compute nodes. The platforms are implemented using a common base structure: Nginx [34] as the HTTPS server connected with FastCGI to a Cppcms [35] backend in C++. We decided to use C++ as the programming language to have a better estimation of the optimal performance that can be achieved with a simple implementation.

---
[6]https://github.com/apastor/ads-chain-cpp-platforms/tree/master/ads-chain-examples

For the purpose of our experiments it is not relevant to use a different key pair for the HTTPS communications and *Ads.chain*. Therefore, we use the same key on each domain for both purposes. The public keys are cached by the *Ads.chain* library on each entity upon the validation of the digital certificate, eliminating the impact that obtaining and validating keys could have on the ad transactions. We issue the digital certificates with the same private Certificate Authority (CA) created with OpenSSL [36] and installed in all the entities as trusted.

For the sample website, we modified a static website template[7] adapting it to the Model-View-Controller pattern of Cppcms to generate dynamic content. We use parameters in the query string to customize the petition to the publisher's web server. This is the mechanism we use to control the number of ad-tags included in the webpage, whether to sign the requests, and to provide a test ID to include in the server-side logs to conduct our performance analysis (See Section V). All the parameters are optional, and the server returns by default one signed ad-tag. As the layout of the ads is not relevant for our purpose, the ad-tags are included as elements of an HTML list that the CSS presents with three elements per row in the lower part of the page. Fig. 4 shows the landing page of the sample website used for the publisher.

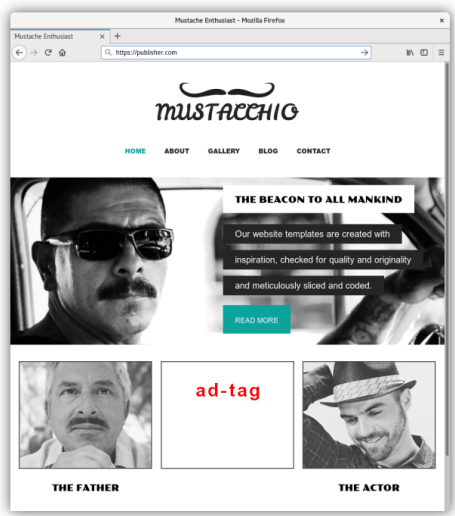

**FIGURE 4.** Image of the landing page of the mock website[7] used for the publisher. The image only contains 1 ad space with ad-tag written in red in the frame.

The programmatic platforms are implemented as Cppcms servers as well. Their first task upon the reception of a request is to check if the fields with signatures are present in the request. If they do, they operate accordingly to the *Ads.chain* specification described in Section III. If not, they process the request following the basic procedure of the platforms without signature verification.

---

[7]The website template is obtained from freewebsitetemplates.com

The auction process in the ad exchange is simulated by launching the bid request to the DSP asynchronously. After 120 ms, the ad URL is extracted from the DSP's bid response and returned to the SSP. Upon responding to the SSP, the ad exchange sends the billing notice to the DSP. The billing notice certifies the DSP as the winner of the auction and includes an updated *Ads.chain* block that delegates the custody of the ad transaction to the DSP.

### B. Ads.chain LIBRARY

The C++ *Ads.chain* library provides classes and functions for the cryptographic operations, network-related functionality, generating UUIDs based on the Unix timestamp, and the logging of execution times. We use this library to implement the protocol in the publisher's web server and programmatic platforms.

The library uses CMake for building the source and Conan [37] for dependency management. The main dependencies of the project are OpenSSL 1.1.1g [36] for the cryptographic operations, RapidJSON [38] for the data structure of the ad transaction information, the Boost Uuid module [39], and Poco [40] for the HTTP requests, caching, and logging. We also use Google's Fruit [41] for dependency injection. In this section, we describe at a high level the design choices we made for the different functionalities implemented in the library to be used by the programmatic platforms. The library is available in a Github repository [9], and the interested reader can refer to the project repository for low-level details.

- **The crypto submodule** of the library provides C++ wrapper classes to OpenSSL. As an entity always signs with the same key, the *Signer* class receives the private key as a parameter in the constructor. The *Verifier* class, as it is expected to verify signatures from different domains, receives the public keys directly in the verify function. The submodule has C++ high-level wrapper classes for the OpenSSL key structures that are especially useful for managing the cache of public keys.

- **The network submodule** includes classes for retrieving and caching the public keys. The public key service caches the keys using the Poco Least Recently Used (LRU) cache with time expiration. If a key is not present, a data access object opens a brief SSL connection to the HTTPS port of the target domain server to retrieve and validate its digital certificate. When a public key is requested to the domain, the public key service adds it to its cache. The network submodule also implements wrapper functions for the HTTPS request and functions for the transformations between query strings and RapidJSON formats. The programmatic platforms can use GET calls when the ad transaction parameters are encoded in the query string (generally on the sell-side) and POST calls when using the OpenRTB JSON objects. Besides, the submodule also has a function to re-create the string that was signed at a given level of the chain of custody. The fields that this function adds to the string are extracted from the *keys-string* field, so that it can be applied to any block in the chain independently of the information signed by the platform generating each specific block.

**- The tools submodule** provides a time-based UUID generator and two stopwatch classes for taking time execution measurements. The UUID generator uses the Boost Uuid format [39] and follows the style of the generators of the library. We implemented it given the lack of a time-based generator in Boost. The first stopwatch is semi-automatic and logs the elapsed time from the object instantiation to the call to the stop function. The other stopwatch is fully automated using RAII (Resource Acquisition Is Initialization) [42] to compute the time between the object construction and destruction. Both stopwatches use the logger passed as an argument to their constructor and allow to set extra fields for additional information of the configuration for which the times are taken.

## V. PERFORMANCE EVALUATION

*Ads.chain* may slightly increase the processing time to render a webpage since it forces to generate the ad transaction ID, create a block, and sign it. Likewise, the processing time of ad transactions in programmatic platforms (SSPs, AdX, and DSPs) may increase with the use of *Ads.chain* due to the need to create a block and its associated digital signature.

As described in Section III-A, the delay incurred by *Ads.chain* must be limited to guarantee a negligible impact in (1) the load time of webpages–since it has been reported that a high page load time affects the user experience [43]– and (2) the overall time required to deliver an ad in the programmatic ecosystem. Increasing the delay associated with the delivery of programmatic ads may reduce the number of ads that are rendered to the user.

In this section, we leverage our lab prototype and the specific functions implemented in the *Ads.chain* library (See Section IV) to evaluate the impact on performance introduced by *Ads.chain* on publishers and programmatic platforms. To this end, we run experiments with and without *Ads.chain*, using two reference digital signature algorithms (See Section II-C2), RSA with a key size of 2048 bits and ECDSA with the curve NIST P-256, and compare the obtained results. First, we analyze how *Ads.chain* affects the publisher's performance, especially on the page serving times. Then, we illustrate the impact on programmatic platforms analyzing the case of the SSP.

An aspect to highlight in our evaluation is that we are assuming that *Ads.chain* executes sequentially to the rest of the actions run by the web server or programmatic platforms, and thus we are reporting marginal delays for a worst-case scenario. In the real world, a server's backend performs multiple tasks in parallel and asynchronously. Therefore, in production environments, *Ads.chain* should be transparent in terms of the delay as backends generally perform other slower operations.

### A. PERFORMANCE AT PUBLISHER's SERVERS

We define the Page Serving Time as the time between the instant the web server receives the query from the browser and the instant when the web server sends the page to the browser. As discussed in Section III, publishers implementing *Ads.chain* include the ad-tags with their signatures on the requested webpage, and signatures are created upon the reception of the request in real-time because they include information linked to the user making the request. Therefore, this metric allows us to objectively measure the impact that *Ads.chain* has on the overall page load time, and thus, on the user experience.

We conduct our evaluation for different web server's capacities. We run stress tests launching the server of our example website on instances of 2, 4, 8, and 16 vcpus from our private OpenStack (See Section IV-A). Moreover, as the signature has to be done for every ad-tag, we conduct tests for various numbers of ad spaces per page, from 1 to 30.[8] We indicate the number of ads and whether to use signed ad-tags with parameters of the query string, as explained in Section IV-A. For launching the requests, we use wrk to generate throughputs from 100 to 1000 requests per second, equivalent to 8.6 M and 86 M requests per day.[9] The duration of the tests is 1 minute, and a script verifies that the server's responses include the ad-tags for correctness. The server execution time for every request is logged with an additional parameter to identify the test. We repeat the battery of tests for the three possible configurations: using signed ad-tags for the two types of keys, RSA and ECDSA, and without signed ad-tags.

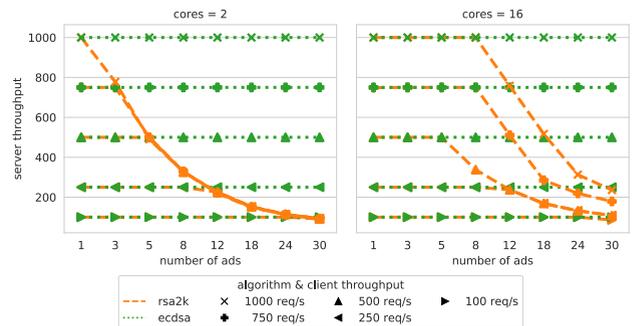

**FIGURE 5.** Throughput achieved by the publisher on the test benchmarks for the smallest and largest server sizes tested. The color and line type indicates the signature algorithm used and the marker the throughput emitted by the client.

Before evaluating the obtained delays, we characterize the impact of implementing the protocol in terms of throughput and resources usage. Fig. 5 shows the throughput achieved in all the test configurations for the publisher servers of 2 and 16 vcpus. Note that we configured a conservative timeout in wrk in order to identify the maximum throughput achievable by the server without errors. The throughput results of wrk show that with RSA, the server could not maintain the rate of the requests for the most demanding configurations. We monitored the CPU and memory usage with pidstat [46]. Fig. 6 shows the maximum CPU usage for each test, including the contribution of the Nginx server and the publisher backend application. The pidstat results show that the CPU overhead with ECDSA is lower than with RSA and that implementing the *Ads.chain* protocol has no impact in terms of memory usage. The results show that the impact on hardware resources is acceptable for the ECDSA keys for all the considered configurations and for the RSA keys for those less demanding scenarios.

---
[8]Previous studies have identified a wide range in the number of ads served per webpage, being 10 the average and 47 the maximum [44].

[9]Data provided by SimilarWeb [45] indicates that, for instance, cnn.com has in average 45M daily page visits.

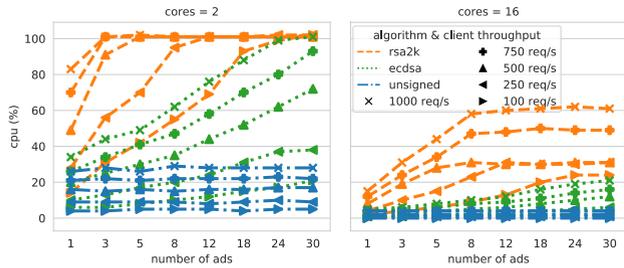

FIGURE 6. CPU usage in the publisher servers of 2 and 16 vcpus for the different tests.

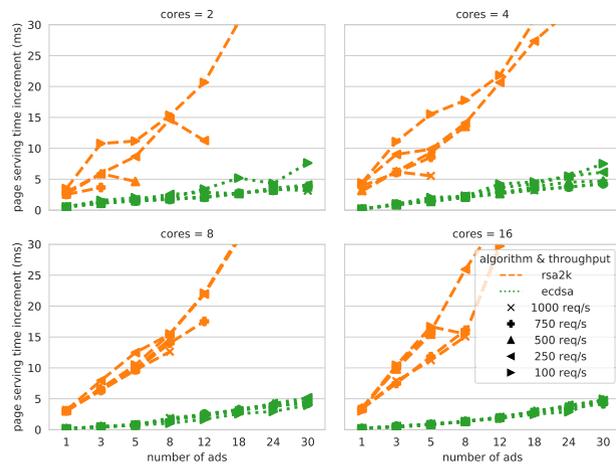

FIGURE 7. Marginal delay's increment on the publisher's page serving time introduced by *Ads.chain* at the 99th percentile.

Next, we focus on analyzing the impact that Ads.chain's signed ad-tags have on the page serving time. For this purpose, we compute the requests times' percentiles for both runs of every test, with and without signed ad-tags. Fig. 7 shows the increment in the page serving times added by *Ads.chain* at the 99th percentile, and Table 1 summarizes the marginal delay per ad space. The results indicate that *Ads.chain* introduces delays ranging from 0.1 ms to 0.6 ms per ad space when the publisher uses ECDSA P-256 keys, depending on the scenario: number of ads per page, server throughput load, and capacity. Instead, when the publisher uses the RSA keys, the additional delay per ad space at the 99th percentile ranges between 0.9 ms and 4.4 ms. The overhead introduced for the generation of UUIDs is in the order of $\mu$ s.

TABLE 1. Marginal delay per ad space introduced at the publisher by *Ads.chain* in the different experiment configurations.

| time (ms) | RSA 2048 | | | ECDSA P-256 | | |
|---|---|---|---|---|---|---|
| | min | mean | max | min | mean | max |
| 90th percentile | 0.8 | 1.7 | 3.3 | 0.1 | 0.2 | 0.4 |
| 95th percentile | 0.8 | 1.9 | 3.9 | 0.1 | 0.2 | 0.4 |
| 99th percentile | 0.9 | 2.4 | 4.4 | 0.1 | 0.2 | 0.6 |

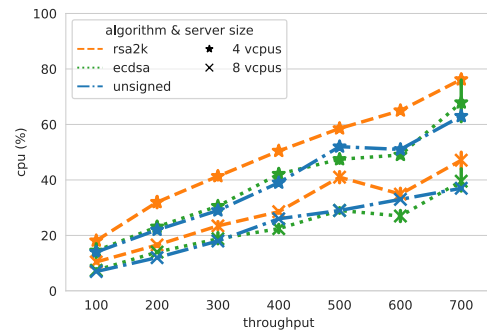

FIGURE 8. Maximum CPU usage at the SSP for the different tests.

Based on these results, we find that for pages with dynamic content, the impact of *Ads.chain* when using ECDSA keys is transparent for the final user as the signatures can be computed in parallel to other tasks. However, using RSA keys may significantly reduce the maximum throughput the server can reach and increment the CPU usage considerably. Therefore, RSA keys have a greater chance of having an impact on the page serving time. The results obtained for ECDSA P-256 on the publisher's signature times are also valid for *Ads.cert* as it uses the same curve.

### B. PERFORMANCE AT PROGRAMMATIC PLATFORMS

To measure the impact of *Ads.chain* in the ad delivery process in programmatic platforms, we consider the case of the SSP in our lab prototype. We characterize this impact by analyzing the overall additional delay and the increment in usage of the server hardware resources. The results for other platforms are expected to be similar to those reported below. We consider the cases where both the publisher and the SSP use the same key type, either RSA 2048 or ECDSA P-256.

We run one experiment per collected metric for each configuration of requests' throughput, server capacity (4 and 8 vcpus),[10] and key type. As in the case of the publisher, we use wrk to generate the requests and pidstat to monitor the resources' usage. The tests duration is 1 minute, and a script checks the responses returned by the server include the ad URL. We deployed the AdX and DSP in servers of 16 vcpus and let the regular flow of ad transaction's

---
[10]This can be considered household commodity hardware. It is expected that stakeholders in the online advertising industry use significantly more powerful hardware infrastructures.

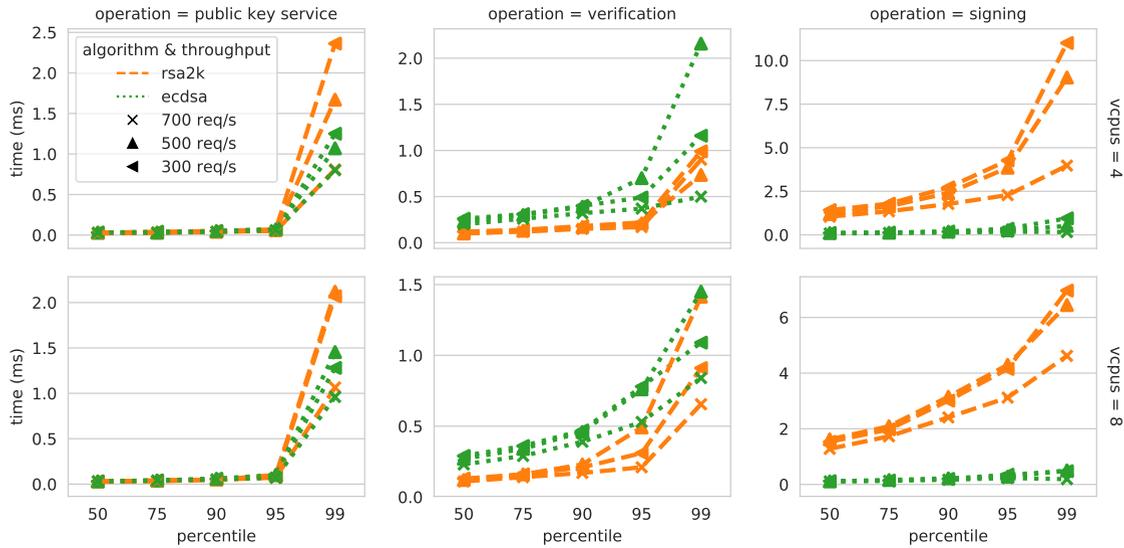

**FIGURE 9.** Processing time at the SSP for the different operations of *Ads.chain*.

messages to happen to replicate the load of a real process in our measurements at the SSP.

We consider different load scenarios ranging between 100 and 700 requests per second, which are equivalent to 8.6 M and 60 M ad requests per day, respectively.[11] The increment on CPU usage when using *Ads.chain* compared to the operation without it is below 15 % in all cases, as shown in Fig. 8. The impact in CPU usage is, in general, not significant with ECDSA and shows more variability with RSA. The error rate measured at the client is below 1.6 % in all the tests, and there is no impact in terms of memory usage.

Then, we computed the delays associated with the different *Ads.chain*'s operations: retrieving the public key of the previous entity (generally from a local cache), verifying the publisher's signature, and signing a new block. Fig. 9 shows the delays for these operations at different percentiles for representative throughputs. Our experiments report better results for the 700 requests/s rate compared to other less demanding rates at the 99th percentile. This indicates that the number of requests with unusually high operation times does not increase with throughput saturation and, instead, depends on the specific load conditions of the server in the specific conducted experiments. Table 2 summarizes the overall delays (considering all the *Ads.chain*'s operations run by the SSP) for the different configurations. When the SSP uses the ECDSA P-256 keys, the overall delay is below 1.4 ms and 4.4 ms at the 95th and 99th percentiles, respectively. With RSA, despite having faster verification times, the slower performance for signing makes the overall delay higher than with ECDSA in all the cases.

Based on these results, we conclude that *Ads.chain* can be implemented by programmatic advertising platforms with a negligible impact on resources usage and delay when using

---

[11] We limit our experiments to 700 requests per second since with the considered servers capacities the throughput returned by the SSP could not match a rate of 800 ad transactions per second.

**TABLE 2.** Overall *Ads.chain* delays at the SSP per ad transaction in the different experiment configurations.

| time (ms) | RSA 2048 | | | ECDSA P-256 | | |
|---|---|---|---|---|---|---|
| | min | mean | max | min | mean | max |
| 90th percentile | 1.9 | 2.9 | 3.7 | 0.5 | 0.7 | 0.8 |
| 95th percentile | 2.2 | 3.9 | 5.1 | 0.5 | 1.0 | 1.4 |
| 99th percentile | 3.7 | 8.7 | 14.4 | 0.8 | 2.6 | 4.4 |

ECDSA keys of the curve NIST P-256. However, using RSA keys at several steps of the ad transaction increases the chances of affecting the overall ad delivery delay.

## VI. CONCLUSION

In this paper we present *Ads.chain*, a protocol that provides end-to-end traceability of individual ad transactions. It offers the required scalability to operate in the current online advertising ecosystem. Moreover, it uses de-facto standard technologies (ad-tags and OpenRTB), guaranteeing an easy and seamless integration in the current programmatic ecosystem.

*Ads.chain* extends the current effort of the IAB in providing traceability in online advertising through the *Ads.txt* and *Ads.cert* solutions. *Ads.chain* addresses the limitations of these techniques and provides (to the best of the authors' knowledge) the first solution meeting the goal pursued by the IAB's efforts to provide end-to-end traceability to ad transactions.

We demonstrate through extensive lab experiments that the impact of *Ads.chain* in the end-user experience browsing webpages and the operation of online advertising intermediaries is expected to be negligible. Our implementation using OpenSSL shows better performance with ECDSA, using the curve NIST P-256 (specified for *Ads.cert*), than with RSA and a key size of 2048 bits. Therefore, although the protocol is independent of the digital signature algorithm used by each entity, we recommend using ECDSA P-256 to ensure there is no additional delay in the ad delivery process.

*Ads.chain* code, as well as the additional code used for its evaluation, are publicly available. We encourage the research community and industry to provide feedback that helps to improve our solution and to conduct further measurements related to its performance. Our current effort focuses on finding interested stakeholders from the online advertising industry to conduct trials in real systems. Additionally, we will work on the definition of scalable auditing systems that automatically analyze the signature chains generated by *Ads.chain* to identify misbehaving entities.

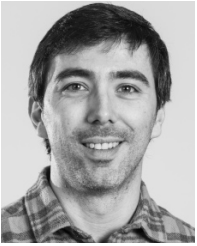
**ANTONIO PASTOR** received the B.Sc. degree in telecommunication technologies engineering and the M.Sc. degree in telematics engineering from the University Carlos III of Madrid, in 2015 and 2016 respectively, where he is currently pursuing the Ph.D. degree in telematics engineering. In 2019, he did an internship with the Brave's Research Group. He has published conference papers at WWW and ACM CoNEXT. His main research interests include applied machine learning, graph theory, and security and privacy.

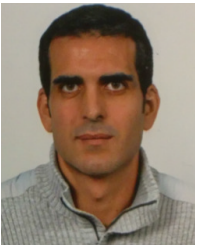
**RUBÉN CUEVAS** received the M.Sc. degree in telecommunications engineering, and the M.Sc. and Ph.D. degrees in telematics engineering from the University Carlos III of Madrid (UC3M), in 2005, 2007, and 2010, respectively, and the M.Sc. degree in network planning and management from Aalborg University, Denmark, in 2006. From January to December 2012, he was a Courtesy Assistant Professor with the Computer and Information Science Department, University of Oregon. He is currently an Associate Professor with the Department of Telematics Engineering, UC3M. He is also the Deputy Director of the UC3M-Santander Big Data Institute (IBiDAT). He has been a PI of 13 research projects funded by the EU H2020, FP7 programs, the National Government of Spain, and private companies. He has overall participated in 27 research projects. He is the coauthor of more than 70 papers in prestigious international journals and conferences, such as ACM CoNEXT, WWW, Usenix Security, ACM HotNets, IEEE Infocom, ACM CHI, IEEE/ACM TON, IEEE TRANSACTIONS ON PARALLEL AND DISTRIBUTED SYSTEMS (TPDS), CACM, PNAS, Nature Scientific Reports, PlosONE or Communications of the ACM. His main research interests include online advertising, web transparency, personalization and privacy, and Internet measurements and its application to solve longstanding problems in other disciplines, such as economy or sociology.

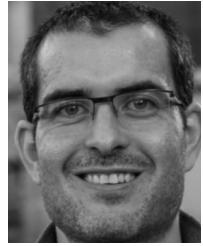
**ÁNGEL CUEVAS** received the B.Sc. degree in telecommunication engineering, and the M.Sc. and Ph.D. degrees in telematics engineering from the Universidad Carlos III de Madrid, in 2006, 2007, and 2011, respectively. He is currently a Ramón y Cajal Fellow (tenure-track Assistant Professor) with the Department of Telematics Engineering, Universidad Carlos III de Madrid, and an Adjunct Professor with the Institut Mines-Telecom SudParis. He is the coauthor of more than 50 papers in prestigious international journals and conferences, such as the IEEE/ACM TRANSACTIONS ON NETWORKING, the ACM *Transactions on Sensor Networks*, the *Computer Networks* (Elsevier), the IEEE NETWORK, the IEEE *Communications Magazine*, WWW, ACM CoNEXT, and ACM CHI. His research interests include Internet measurements, web transparency, privacy, and P2P networks. He was a recipient of the Best Paper Award from the ACM MSWiM 2010.

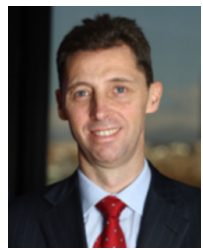
**ARTURO AZCORRA** (Senior Member, IEEE) received the M.Sc. degree in telecommunications engineering and the Ph.D. degree from the Universidad Politécnica de Madrid, in 1986 and 1989, respectively, and the M.B.A. degree from the Instituto de Empresa, in 1993. He has a double appointment as a Full Professor (with Chair) with the Department of Telematics Engineering, University Carlos III of Madrid, and as the Director of the IMDEA Networks. He has coordinated the CONTENT and E-NEXT European Networks of Excellence. He has published over 100 scientific papers in books, international journals, and conferences. He has served as a Program Committee Member in numerous international conferences.

∙ ∙ ∙